\documentclass[10pt, conference]{IEEEtran}
\IEEEoverridecommandlockouts

\usepackage{amsmath,amssymb,amsfonts}
\usepackage{algorithm}
\usepackage{algpseudocode}
\usepackage{amsmath}
\usepackage{textcomp}
\usepackage{xcolor}
\usepackage{comment}
\usepackage{graphicx}
\usepackage{svg}
\svgsetup{inkscapelatex=false}
\usepackage{array}
\usepackage{booktabs}
\usepackage{makecell}
\usepackage{multirow}
\usepackage{threeparttable}
\usepackage{soul} 
\def\BibTeX{{\rm B\kern-.05em{\sc i\kern-.025em b}\kern-.08em
    T\kern-.1667em\lower.7ex\hbox{E}\kern-.125emX}}

\usepackage[backend=biber,style=ieee]{biblatex}
\newcommand{\verticalcell}[2][2.3cm]{%
  \rotatebox[origin=c]{90}{\parbox[c]{#1}{\centering\bfseries #2}}%
}

\newcolumntype{L}[1]{>{\raggedright\arraybackslash}m{#1}}
\newcolumntype{C}[1]{>{\centering\arraybackslash}m{#1}}

\algrenewcommand\algorithmicrequire{\textbf{Input:}}
\algrenewcommand\algorithmicensure{\textbf{Output:}}
\algrenewcommand\algorithmiccomment[1]{\hfill{\footnotesize$\triangleright$~#1}}

\begin{document}

\title{Modeling and Recovering Hierarchical Structural Architectures of ROS 2 Systems from Code and Launch Configurations using LLM-based Agents\\
}

\author{\IEEEauthorblockN{
Mohamed Benchat (0000-0002-2577-9925),
Dominique Briechle (0009-0000-2610-3399),\\
Raj Chanchad (0009-0001-1051-9303), 
Mitbhai Chauhan (0009-0008-2555-1198),\\
Meet Chavda (0009-0007-3038-9597),
Ruidi He (0009-0006-3849-7659),\\
Dhruv Jajadiya (0009-0004-4142-6179),
Dhruv Kapadiya (0009-0005-2759-9812),\\
Nidhiben Kaswala (0009-0006-1162-5913),
Daniel Osterholz (0009-0007-2940-5473),\\
Andreas Rausch (0000-0002-6850-6409),
Meng Zhang (0000-0002-9831-9356)
}

\IEEEauthorblockA{
Institute for Software and Systems Engineering\\
Clausthal University of Technology\\
Clausthal-Zellerfeld 38678, Germany\\
}
\thanks{\textasteriskcentered\ Authors are listed alphabetically; author order does not indicate contribution.}
}

\maketitle

\begin{abstract}
Model-Driven Engineering (MDE) relies on explicit architecture models to document and evolve systems across abstraction levels. For ROS~2, subsystem structure is often encoded implicitly in distributed configuration artifacts---most notably launch files---making hierarchical structural decomposition hard to capture and maintain. Existing ROS~2 modeling approaches cover node-level entities and wiring, but do not make hierarchical structural (de-)composition a first-class architectural view independent of launch artifacts.

We contribute (1) a UML-based modeling concept for hierarchical structural architectures of ROS~2 systems and (2) a blueprint-guided automated recovery pipeline that reconstructs such models from code and configuration artifacts by combining deterministic extraction with LLM-based agents. The ROS~2 architectural blueprint (nodes, topics, interfaces, launch-induced wiring) is encoded as structural contracts to constrain synthesis and enable deterministic validation, improving reliability.

We evaluate the approach on three ROS~2 repositories, including an industrial-scale code subset. Results show high precision across abstraction levels, while subsystem-level recall drops with repository complexity due to implicit launch semantics, making high-level recovery the remaining challenge.
\end{abstract}


\section{Introduction}
Model-Driven Engineering (MDE) promotes explicit, tool-supported models as primary artifacts to communicate, analyze, and evolve software architectures \cite{schmidt}. Such models underpin architecture documentation by capturing stable abstractions beyond implementation details. Established templates such as C4 \cite{brown2013software} and arc42 \cite{starke2019arc42} structure architectural descriptions along hierarchical levels and complementary views. In particular, the component- and code-oriented perspectives (C4 L3/L4) require an explicit representation of hierarchical structural (de-)composition \cite{brown2013software}.

We study this challenge for ROS~2 systems, where subsystem structure is often encoded implicitly in launch-based wiring rather than as explicit architecture models. Although ROS-specific modeling approaches exist (e.g. \cite{10.1145/3763169, 7416545, 7926539, Winiarski_2023, Wanninger2021}), they primarily capture node-level concepts and communication and do not adequately support hierarchical structural composition across abstraction levels. 

Moreover, maintainability is a practical obstacle: long-lived systems evolve continuously, and documented architectures drift from implementation, eroding the value of documentation for onboarding, shared understanding, and safe evolution \cite{koschke2005rekonstruktion}. For ROS~2, this risk is amplified because key architectural decisions are dispersed across heterogeneous artifacts, in particular source code, launch files, and parameter configurations.

Recent work explores LLM-based approaches for architecture recovery and documentation generation (e.g., \cite{semarch}), but generic, unconstrained inference can yield abstractions that are incomplete, inconsistent, or hard to validate. We therefore leverage system-specific architectural knowledge to constrain recovery: since the target is a ROS~2 system, the admissible element types and relations are largely known (nodes, topics, services, namespaces, and launch-induced wiring). We encode this ROS~2 blueprint as constraints and contracts that bound LLM-based synthesis and enable deterministic validation.

Concretely, this paper contributes (1) a UML-based modeling concept for hierarchical structural architectures of ROS~2 systems, (2) a blueprint-guided automated recovery system combining deterministic extraction and LLM-based agents to reconstruct such models from code and launch configurations, and (3) an empirical demonstration on representative ROS~2 systems. The remainder is structured as follows: Section~2 provides background and related work. Section~3 introduces the modeling concept. Section~4 presents the recovery system. Section~5 reports the evaluation. and Section~6 concludes with limitations and future work.

\section{Related Work}
\subsection{Modeling/description techniques for ROS~2 architectures} 
ROS~2 follows a loosely coupled, component-oriented paradigm in which nodes interact via publish--subscribe topics, services, and actions, while system integration is specified via build artifacts (e.g., CMake), launch files, and parameter configurations. Architectural understanding therefore requires consolidating implementation-level wiring and abstracting it into stable subsystem structures across artifacts. Several approaches model ROS-based architectures, but differ in their support for ROS~2 concepts, launch artifacts, and higher-level structural decomposition. MeROS provides an up-to-date ROS~1/ROS~2 SysML/UML-based metamodel organized into structural and behavioral views \cite{Winiarski_2023}, and notes the effort and synchronization challenges of manual UML/SysML modeling, motivating automated extraction \cite{Winiarski_2023}.

ROSpec consolidates scattered node connections via a dedicated DSL \cite{10.1145/3763169}. ROSMOD supports model-driven, component-based development \cite{7416545}. AADL enables formal real-time and safety analysis \cite{7926539}, and ROSSi offers graphical launch editing, though limited to ROS~1-level functionality \cite{Wanninger2021}.

Overall, existing approaches model ROS concepts nodes, configuration, or tooling effectively, but they do not represent hierarchical structural (de-)composition as an explicit architectural view independent of launch artifacts \cite{Wanninger2021}. To the best of our knowledge, no approach simultaneously treats hierarchical structural composition as a first-class view while keeping models consistent with evolving code and launch-induced wiring at reasonable engineering effort, motivating our UML-based modeling concept and blueprint-guided approach.

\subsection{Structural architecture recovery and LLM-based recovery}
Static architecture recovery traditionally relies on reverse engineering and structural analysis to extract elements and dependencies from code, but results often remain low-level or incomplete with respect to architecturally meaningful abstractions. Recent LLM-based work attempts to bridge this gap by using LLMs to abstract and summarize structure, for example, by converting low-level representations into higher-level components and optionally producing complementary behavioral views \cite{hatahet2025generatingsoftwarearchitecturedescription}. Yet, broader results remain inconsistent: code-only input can be insufficient for robust architectural inference compared to natural-language architecture documentation, and recovered views may miss key architectural decisions when they are distributed across heterogeneous artifacts \cite{hatahet2025generatingsoftwarearchitecturedescription, fuchs2025enabling}.

LLM-based recovery without domain constraints risks generating plausible yet unsupported architectural abstractions, particularly in settings such as ROS~2 where architectural semantics are distributed across source, build, and launch artifacts. Prior work shows that abstraction accuracy improves when LLMs are enriched with domain-specific knowledge (e.g., via structured prompting or few-shot examples) and when recovery is guided by an explicit definition of architectural elements in the target domain~\cite{hatahet2025generatingsoftwarearchitecturedescription}.

Following this principle, our approach adopts a blueprint-guided strategy tailored to ROS~2. We explicitly exploit the fact that ROS~2 systems follow a recognizable architectural blueprint comprising nodes, topics, services/actions, namespaces, and launch-induced wiring. This blueprint is embedded into the recovery pipeline as extraction contracts, structured prompts, and validation rules, thereby constraining the LLM agents to produce abstractions aligned with the intended architectural viewpoint. In contrast to generic pattern-library guidance, these constraints are domain-specific and derived directly from the ROS~2 blueprint, enabling deterministic validation against explicit structural contracts and reducing ambiguity in the recovered models.

\section{UML-based Modeling for Hierarchical Structural Architecture of ROS 2 Systems}
\label{chapter:uml_profile}

To represent hierarchical structural decomposition of ROS~2 systems, we define a UML-based modeling concept that maps ROS~2 source code and launch/configuration artifacts to architectural elements. It specifies admissible classifiers, ports, interfaces, and composition rules to model node definitions and interfaces, instantiation/configuration, namespace propagation/remapping, and launch-induced subsystem composition.

We introduce four core concepts: \emph{AtomicRosNodeClassifier}, \emph{RosNodePart}, \emph{ComposedRosNodeClassifier}, and \emph{Namespace}. \emph{AtomicRosNodeClassifier} captures code-level building blocks (C4 L4) derived from ROS~2 node classes, while \emph{ComposedRosNodeClassifier} captures launch-induced subsystem composition (C4 L3). We first define code-level classifiers and port semantics, then instantiation and launch-induced composition, and finally namespace scoping rules for hierarchical runtime names.

\subsection{AtomicRosNodeClassifier}
\begin{figure*}[htbp]
\centerline{\includegraphics[width=0.95\textwidth]{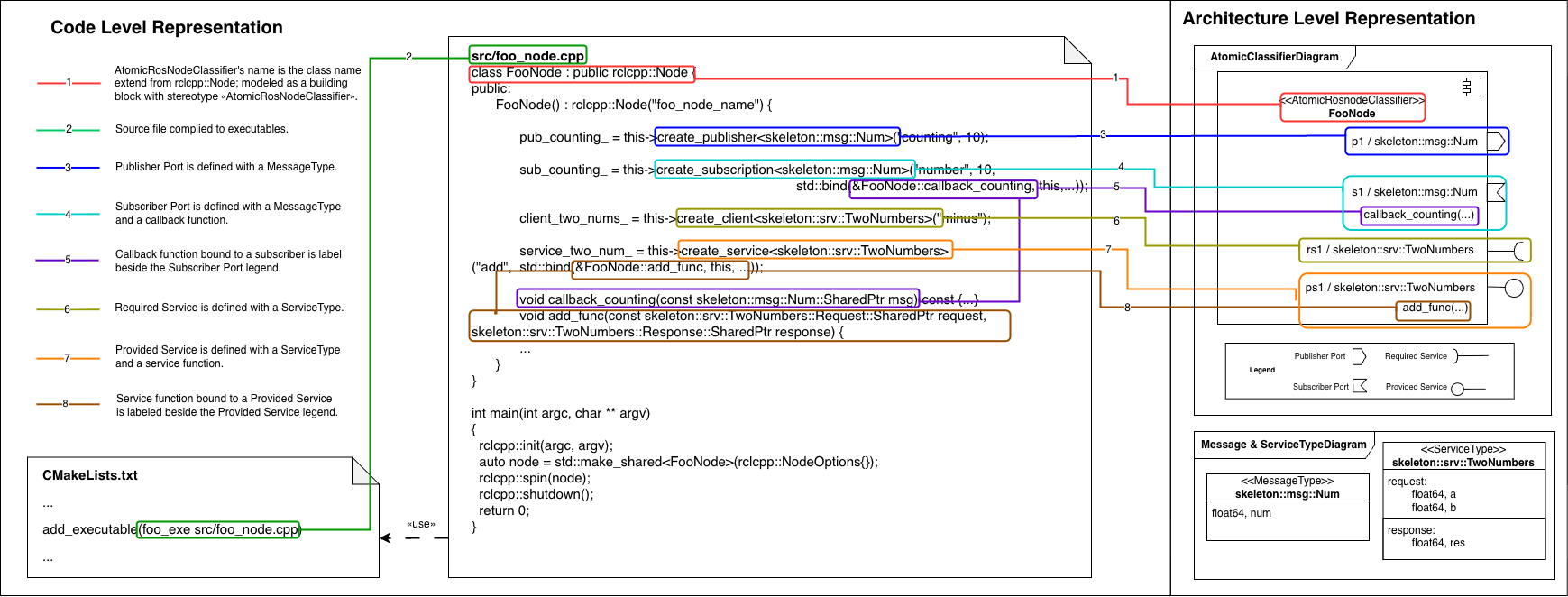}}
\caption{Code-level and Architecture-level representation of the \emph{AtomicRosNodeClassifier} \texttt{FooNode}, including the \emph{AtomicClassifierDiagram} and the associated \emph{Message \& ServiceTypeDiagram}. The colored Trace lines indicate the mapping between model annotations and code-level identifiers.}
\label{fig:atomic-classifier}
\end{figure*}

\begin{figure*}[htbp]
\centerline{\includegraphics[width=0.95\textwidth]{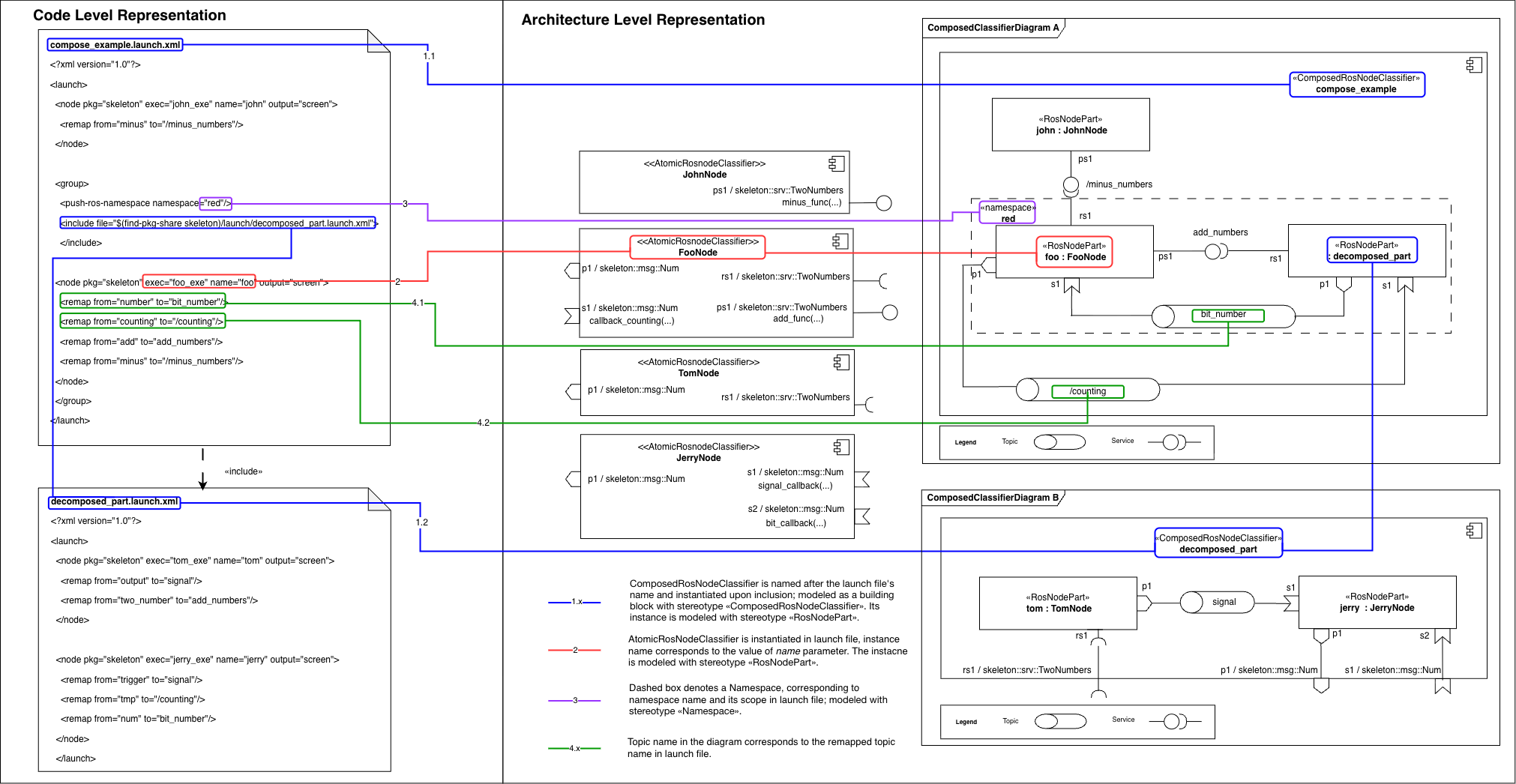}}
\caption{Code- and architecture-level representation of the \emph{ComposedRosNodeClassifier} \texttt{Compose\_example}. The architecture-level view comprises two \emph{ComposedClassifierDiagram}s; atomic classifier structures are embedded but not explicitly represented as separate diagrams. Colored Trace lines indicate the mapping between model annotations and code-level identifiers.}
\label{fig:composed-classifier}
\end{figure*}

The \textbf{\emph{AtomicRosNodeClassifier}} constitutes the fundamental structural unit of our modeling concept, representing the smallest explicitly modeled architectural element in ROS~2 systems.

\paragraph{Mapping to Source Code and Naming}
It corresponds to a ROS~2 node class inheriting from \texttt{rclcpp::Node}, and its name is identical to the source-level class name. As shown in Fig.~\ref{fig:atomic-classifier} Trace 1, the class \texttt{FooNode} extends \texttt{rclcpp::Node} and is modeled as an \emph{AtomicRosNodeClassifier} named \texttt{FooNode}. Trace 2 in Fig.~\ref{fig:atomic-classifier} additionally indicates that the source class is integrated into the build configuration and compiled into an executable artifact referenced in launch file.

\paragraph{Port Definition and Typing Semantics}

Within the class body, ports are defined for topic- and service-based communication. The four kinds of ports (publisher/subscriber port and required/provided service) are each represented by distinct symbols in the Fig.~\ref{fig:atomic-classifier}. As shown in Trace 3 in Fig.~\ref{fig:atomic-classifier}, a publisher port defined via \texttt{create\_publisher} is parameterized by the MessageType \texttt{skeleton::msg::Num}. This MessageType is explicitly captured in the corresponding legend in the diagram. Traces 4, 6, and 7 in Fig.~\ref{fig:atomic-classifier} illustrate analogous mappings for the remaining port types. Each port carries a unique identifier within its enclosing classifier, following the format \texttt{identifier / type} (e.g., \texttt{p1 / skeleton::msg::Num} for a publisher port). Furthermore, as shown in Traces 5 and 8 in Fig.~\ref{fig:atomic-classifier}, subscriber ports and provided services additionally indicate the bound callback function in the diagram, annotated by the function name.

\paragraph{Interface Types}
The data contracts carried by ports are represented by two reusable type classifiers.
A MessageType models a ROS~2 message definition (\texttt{.msg} file) as a set of typed fields and serves as the type of publisher and subscriber ports. A ServiceType models a ROS~2 service definition (\texttt{.srv} file) with request and response sections and serves as the type of required and provided service.

\subsection{RosNodePart}

Each \textbf{\emph{RosNodePart}} is an instance of either a \emph{ComposedRosNodeClassifier} or an \emph{AtomicRosNodeClassifier}.

\paragraph{Instantiation and Structural Conformance}

The naming of a \emph{RosNodePart} follows the UML-style notation:
\texttt{\textit{Name} : \textit{typeName}}.
The typeName is derived from the corresponding \emph{AtomicRosNodeClassifier} or \emph{ComposedRosNodeClassifier}.

For a \emph{RosNodePart} instantiated from an \emph{AtomicRosNodeClassifier} \texttt{FooNode} in Fig.~\ref{fig:composed-classifier} Trace 2, the Name \texttt{foo} is obtained from the \texttt{name} parameter specified in the launch file, which overwrites node name defined in source file. For a \emph{RosNodePart} instantiated from a \emph{ComposedRosNodeClassifier} \texttt{FooNode} in Fig.~\ref{fig:composed-classifier} Trace 1.1, the \textit{Name} is anonymous by default and may be customized according to project-specific conventions in Trace 1.2.

Since a \emph{RosNodePart} is instantiated from either an \emph{AtomicRosNodeClassifier} or a \emph{ComposedRosNodeClassifier}, it shall conform structurally to its defining classifier.
Accordingly, the ports—including their kind, and associated \emph{MessageTypes} or \emph{ServiceTypes}—must correspond exactly to those defined in the classifier specification. Moreover, each port of a \emph{RosNodePart} preserves the identifier defined for the corresponding port in its classifier.

\paragraph{Interfaces Between RosNodeParts}\label{par:interface-rosnodeparts}

\emph{RosNodePart}s communicate via topics by connecting publisher and subscriber ports. A topic is labeled with its name after launch-level remapping (cf. Fig.~\ref{fig:composed-classifier}). As shown in Fig.~\ref{fig:composed-classifier} Trace~4.1, the source-level topic \texttt{number} defined in class \texttt{FooNode} is remapped to \texttt{bit\_number}, which is the name annotated in the diagram. A topic connects exclusively to publisher and subscriber ports. Any number of such ports may attach to the same topic, provided they share an identical MessageType, which determines the topic’s type. \emph{RosNodePart}s communicate via services by connecting provided service and required service. A service is labeled with its name after remapping (cf. Fig.~\ref{fig:composed-classifier}).
A service connects only to servers and clients. Exactly one server and any number of clients may attach to a service, all sharing the same ServiceType.

\subsection{ComposedRosNodeClassifier}
A \textbf{\emph{ComposedRosNodeClassifier}} represents the composition of multiple \emph{RosNodePart}s.

\paragraph{Mapping to Launch Files and Naming}

Each \emph{ComposedRosNodeClassifier} corresponds to a launch file, and its name is derived from the launch file name in Fig.~\ref{fig:composed-classifier}, Trace~1.1. Since ROS~2 launch files may include other launch files, forming hierarchical configuration dependencies, this structure is preserved in the model: a \emph{ComposedRosNodeClassifier} may be instantiated in another \emph{ComposedRosNodeClassifier}, thereby capturing launch-level inclusion as hierarchical composition.

\paragraph{Port Definition and Typing Semantics}\label{par:port-def-type-semant}

Like an \emph{AtomicRosNodeClassifier}, a \emph{ComposedRosNodeClassifier} may define ports.
If a port of an internal \emph{RosNodePart} is not connected to another internal part within the same \emph{ComposedRosNodeClassifier}, it is treated as an external port of the composed classifier and annotated using the same \texttt{identifier / MessageType} format. The same rule applies to service servers and clients.

\subsection{Namespace}
Dashed boxes in the diagram represent namespaces defined in launch files in Fig.~\ref{fig:composed-classifier} Trace~3.

\paragraph*{Namespace Scope and Name Recovery}

All \emph{RosNodePart}s, topics, and services enclosed within a namespace fall under its scope, and their hierarchical runtime names are prefixed accordingly. As shown in Trace~4.1, the topic \texttt{bit\_number} is affected by the namespace \texttt{red} and is therefore placed inside the namespace box. 

In contrast, in Trace~4.2, the topic \texttt{/counting} remains outside the namespace scope, since names beginning with \texttt{/} are absolute and not subject to prefixing. Namespace scope propagates hierarchically. If a \emph{RosNodePart} instantiated from a \emph{ComposedRosNodeClassifier} is placed within a namespace, the namespace applies recursively to all nested \emph{RosNodePart}s, topics, and services on their names.

\section{Blueprint-Guided Automated Architecture Recovery System} 
\label{section:recoveryarch}
\subsection{Overall Concept}
\label{section:overall_concept}
Architecture recovery in this work targets a well-defined representation of hierarchical structural ROS~2 systems. The previously introduced UML-based modeling concept specifies admissible architectural entities — nodes, communication ports, namespaces, and hierarchical subsystems — together with their structural constraints. Recovery is thus formulated as reconstructing these elements from heterogeneous repository artifacts (e.g., source code, build files, launch files, package manifests) such that the resulting model conforms to the profile. 

ROS~2 repositories exhibit a structural separation between a definition phase and an integration phase \cite{10.1145/3763169}. In the definition phase, atomic nodes are implemented as classes declaring publishers, subscribers, services, and callbacks. In the integration phase, launch and build artifacts establish runtime composition by instantiating executables, assigning namespaces, applying remappings, and orchestrating multi-node subsystems. This separation implicitly encodes architectural hierarchy: node implementations serve as stable primitives, whereas higher-level subsystems emerge through distributed configuration semantics.

Recovery must reflect this decomposition. Atomic nodes can be identified deterministically from source artifacts because node identity is syntactically encoded (e.g., inheritance from \texttt{rclcpp::Node}), enabling rule-based static analysis. In contrast, a hierarchical system structure is distributed across launch files, manifests, and build configurations and cannot be reconstructed by structural parsing alone.

These characteristics motivate a hybrid recovery strategy that combines deterministic structural extraction with constrained semantic synthesis to reconstruct distributed integration semantics such as subsystem boundaries and orchestration relations. Rather than assuming generic architectural patterns \cite{zhao2026softwarear}, our approach relies on a ROS-specific architectural modeling concept based on UML. This profile specifies admissible elements (e.g., nodes, ports, namespaces), their structural relations, and composition constraints derived from ROS~2 repository conventions. It defines a structurally constrained modeling of ROS~2 systems, but does not yet constitute a fully formal semantic metamodel. The blueprint thus constrains LLM-assisted synthesis, ensuring that reconstructed architectures remain traceable to repository artifacts and structurally consistent with ROS~2 conventions.

\subsection{High-Level Architecture of the Automated Recovery System}
\begin{figure}[t]
\centering
\includegraphics[width=\columnwidth]{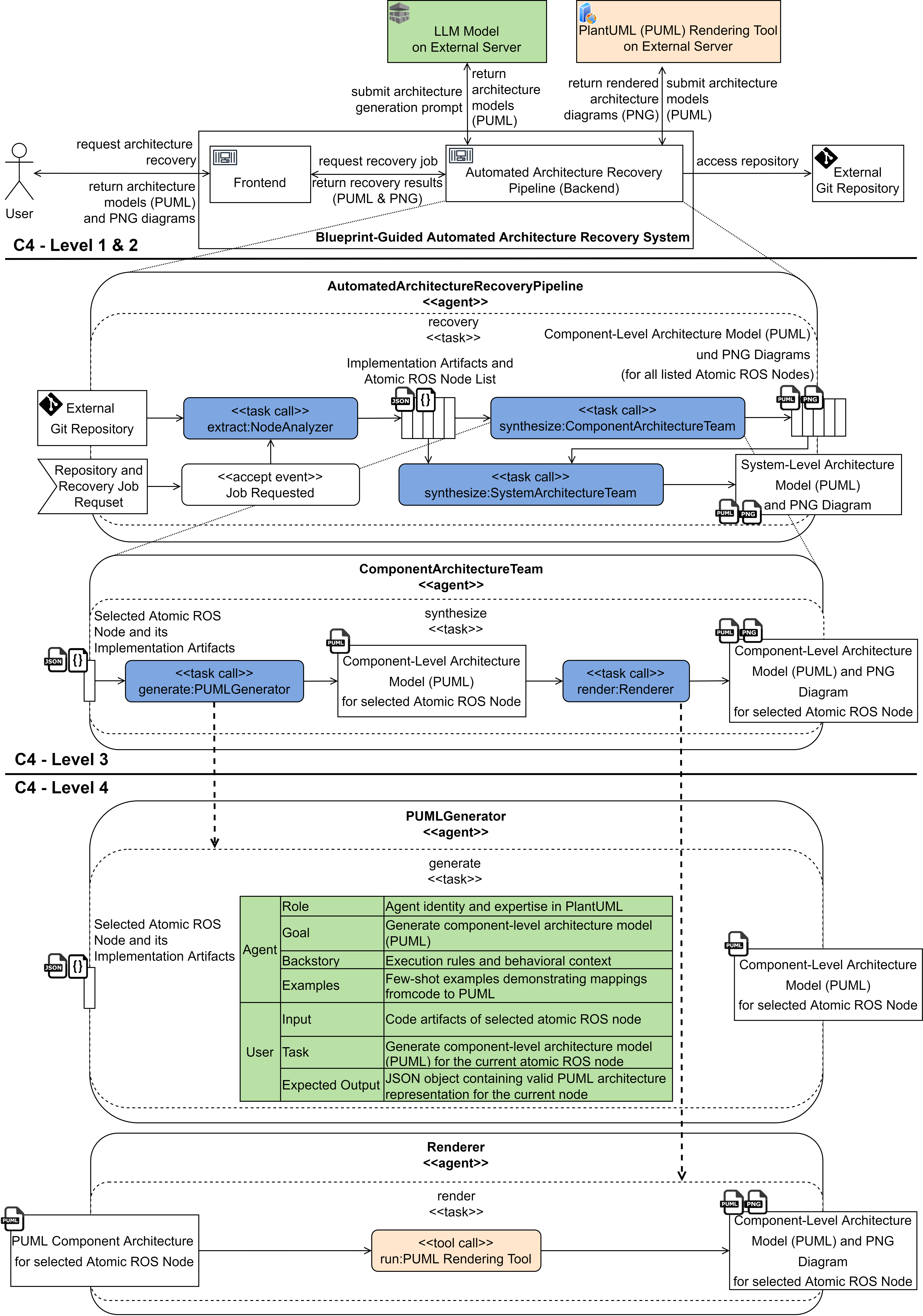}
\caption{C4-conformed Architecture of Blueprint-Guided Automated Architecture Recovery System (retrofitted after \cite{brown2013software}).}
\label{fig:pipeline_architecture}
\end{figure}

Based on these principles, a blueprint-guided automated architecture recovery system is developed (Fig.~\ref{fig:pipeline_architecture}). The architecture is described using an adapted C4 model \cite{brown2013software} that explicitly represents AI agents, inference services, and tool interactions.

The system consists of a \textit{frontend} and a \textit{backend}. The \textit{frontend} accepts a recovery request and returns generated architectural artifacts in textual (PUML) and rendered (PNG) formats. The user's recovery request triggers \textit{backend} processing via a job submission event.

The backend orchestrates a CrewAI-based \textit{Automated Architecture Recovery Pipeline} interacting with external repositories, LLM inference services, and PlantUML rendering \cite{crewai2024introduction}.

The pipeline is structured to reflect the hybrid recovery strategy, explicitly separating deterministic static analysis from AI-assisted semantic architecture synthesis. Three principal agents are defined: a \textit{NodeAnalyzer} performing deterministic extraction of atomic ROS nodes from implementation artifacts, and two agents forming the AI-assisted synthesis layer, namely a \textit{ComponentArchitectureTeam} for component-level abstraction and a \textit{SystemArchitectureTeam} for system-level reconstruction. This strict separation prevents semantic contamination of syntactic extraction and preserves reproducibility of architecture recovery.

\subsubsection {Deterministic Analysis for Extracting Atomic ROS Nodes}
\label{subsection:deterministicanaly}
The \textit{NodeAnalyzer} performs rule-based static analysis on source and header files and produces a structured JSON artifact containing atomic ROS node identities and corresponding file paths. This artifact constitutes the defined interface between structural extraction and semantic synthesis. Because extraction relies solely on explicit syntactic markers, it is fully deterministic and reproducible.

\subsubsection{AI-assisted Synthesis for Architecture Recovery}
AI-assisted architecture synthesis operates at two abstraction levels.

At the component level, the \textit{ComponentArchitectureTeam} consumes the JSON artifact and processes each atomic ROS node sequentially. For each node, associated implementation artifacts are analyzed, and a dedicated \textit{PUMLGenerator} agent communicates with an external LLM inference service to synthesize a PlantUML component-level model. The prompt structure, summarized in tabular form in Fig.~\ref{fig:pipeline_architecture}, encodes architectural abstraction rules and schema constraints derived from a predefined blueprint. Inference is therefore not unconstrained text generation but blueprint-bounded synthesis. Few-shot prompting reinforces these structural constraints through external, system-independent examples that illustrate the modeling format and admissible architectural constructs without containing repository-specific information about the recovered system.

The agent produces a structured JSON artifact containing syntactically valid PlantUML component-level models. Subsequently, a dedicated \textit{Renderer} agent transforms these PlantUML specifications into PNG-based visual diagrams. Each generated component-level architecture model (PUML) preserves traceability to its originating implementation artifacts (e.g., source and header files) and constitutes the lowest abstraction layer of the reconstructed architecture model.

At the system level, the \textit{SystemArchitectureTeam} reconstructs hierarchical runtime composition. System structure emerges from heterogeneous configuration artifacts, including launch files, build configurations, package manifests, and parameter specifications. The agent consumes (1) the JSON artifact of extracted nodes, (2) repository configuration artifacts, and (3) generated component-level architecture models, integrating structural identities, configuration semantics, and abstracted component views into a unified system model. Through blueprint-guided reasoning, it synthesizes executable-to-node mappings, namespace hierarchies, subsystem boundaries, and inter-node communication relations. Another \textit{PUMLGenerator} agent produces here the corresponding system-level model (PUML) under schema-constrained prompting, and another \textit{Renderer} agent generates the final system architecture diagram. Together with the component-level architecture models, they form a coherent multi-level set of architectural models derived from heterogeneous implementation and configuration artifacts under controlled semantic constraints.

\section{Experimental Evaluation}
\subsection{Evaluation Setup with Case Studies}
\begin{figure}[t]
\includegraphics[width=\columnwidth]{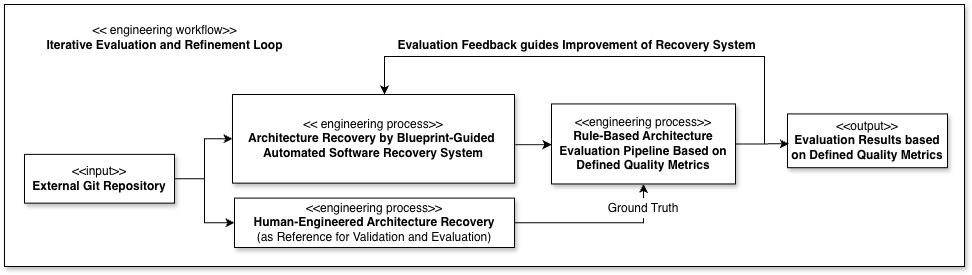}
\caption{Setup of Experimental Evaluation Pipeline}
\label{fig:processflow-eval-val}
\end{figure}

Our evaluation validates the structural correctness of automatically generated \emph{Component-} and \emph{System-level Architecture Models} against manually engineered references using a rule-based deterministic validation pipeline (cf. Fig.~\ref{fig:processflow-eval-val}). Evaluation operates directly on PlantUML artifacts to ensure layout independence. Both models are normalized into internal representations capturing semantically relevant elements—node identities, communication interfaces (topics and services), message/service types, namespace bindings, and directed connectivity—while excluding rendering attributes. Models are accepted only if UML-conformant and traceable; equivalence is assessed at the level of communication semantics rather than syntax.

\textbf{Case Study I – Controlled Synthetic Example without Launch Files:}  
A synthetic ROS~2 system (665 LOC; 6 node classes; 3 launch files) with manually engineered references\footnote{\url{https://github.com/Ruidi345/Controlled_Synthetic_Example}} is used. Although the repository contains launch files, launch artifacts are explicitly excluded from the recovery process. Architecture reconstruction relies solely on source-level node class definitions. The system is structurally minimal and contains limited functional logic, focusing primarily on explicit atomic components.

\textbf{Case Study II - Controlled Synthetic Example with Launch Files:}  
Using the same repository as Case Study I, launch artifacts are incorporated, including nested inclusion, namespace scoping, and runtime name resolution. Only 3 of the 6 node classes are instantiated via the 3 launch files. The expected output includes component- and system-level models reflecting the effective instantiation set, hierarchical composition, namespace propagation, and launch-induced communication.

\textbf{Case Study III - Industrial-Scale Autoware Subset:}  
A larger ROS~2 repository ($\approx$14{,}000 LOC; 3 node classes; 2 launch files)\footnote{https://github.com/Ruidi345/Industrial-Scale\_Autoware\_Subset} is used to assess scalability. The repository preserves a coherent Autoware-subsystem while excluding unrelated components. The expected output is a multi-level architectural model capturing recovered components and hierarchical composition consistent with launch orchestration.

For all studies, ground-truth architectures are derived through systematic inspection of implementation and configuration artifacts by ROS~2 domain experts. Both reference models and automatically generated results are available in the corresponding repository under the \texttt{docs} folders.

\subsection{Architecture Evaluation Pipeline with Implemented Metrics}
Recovered component- and system-level PUML models are evaluated against expert-engineered references using a rule-based deterministic pipeline (cf.Fig.\ref{fig:processflow-eval-val}), with metrics targeting semantic architectural conformance rather than structural similarity.

Clustering-based recovery measures such as MojoFM, ARI, C2C${cvg}$, A2A, or A2A${adj}$ \cite{zhao2026softwarear}, as well as mistake-detection, heuristic, and learning-based similarity methods \cite{10.1145/3550356.3561583,10.1145/3417990.3418741,SONG2024123927,yuan2020structural}, are unsuitable for ROS~2 architectures. These approaches evaluate either clustering similarity or global graph resemblance but do not validate typed communication semantics, port roles, namespace propagation, or launch-induced remapping as defined in our modeling concept.

Architectural correctness is defined as conformance to the UML-based modeling concept (cf.Section\ref{chapter:uml_profile}); evaluation therefore targets profile-defined elements. The metrics in TableI capture architectural properties of atomic ROS nodes and their compositions. Component-level models represent atomic nodes, while system-level models represent their compositions (cf.Fig.\ref{fig:atomic-classifier}, Fig.\ref{fig:composed-classifier}); metrics are assessed independently at both levels.

After parsing, generated and reference models are transformed into canonicalized sets of metric elements encoded as structured attribute tuples to ensure deterministic, order-independent comparison.

Following established practice in automated model extraction and reconstruction research \cite{yang2022towardsautouml,Chen2023AutomatedDomainModeling}, equality is defined as exact encoding matches. Elements shared by both models are True Positives (TP), those only in the reference are False Negatives (FN), and those only in the generated model are False Positives (FP); True Negatives are undefined due to the unbounded element space.

Despite blueprint-specific metric elements, we employ Precision/Recall/F1 as the standard paradigm for element-level model reconstruction, ensuring statistical comparability while accommodating domain-specific element definitions.

For each PUML model, TP, FP, and FN are obtained by counting matching, additional, and missing elements. Precision, recall, and F1-score are computed as

\[
P = \frac{TP}{TP + FP}, \quad
R = \frac{TP}{TP + FN}, \quad
F1 = \frac{2PR}{P + R}.
\]

Scores are computed per model and macro-averaged across all evaluated component- and system-level models. Precision quantifies structural correctness (penalizing False Positives, i.e., hallucinated elements), recall quantifies structural completeness (penalizing False Negatives, i.e., missing elements), and F1 balances both aspects to measure profile-constrained architectural conformance.

\renewcommand{\arraystretch}{1.1}
\begin{table}[t]
\centering
\footnotesize
\setlength{\tabcolsep}{3pt}
\caption{Implemented Metrics Matching to the UML-Based Modeling Concept (in Section \ref{chapter:uml_profile})}
\label{tab:metrics-impl}

\begin{tabular}{m{1.0cm} m{5.0cm} m{2.0cm}}
\toprule
\textbf{Level} & \textbf{Metric Element} & \textbf{Modeling Concept}\\
\midrule
\multirow{6}{*}[1em]{\verticalcell[2.55cm]{\textbf{AtomicClassifier-Diagram}\\(cf. Fig. \ref{fig:atomic-classifier})}}
& Name of AtomicRosNodeClassifier & Trace 1\\
\cline{2-3}
& Stereotype of AtomicRosNodeClassifier  & Trace 1\\
\cline{2-3}
& MessageType & Traces 3 \& 4\\
\cline{2-3}
& Name of Callback Function & Trace 5\\
\cline{2-3}
& ServiceType & Traces 6 \& 7\\
\cline{2-3}
& Name of Service Function & Trace 8\\
\midrule
\multirow{7}{*}[+0.8em]{\verticalcell[3cm]{\textbf{ComposedClassifier-Diagram}\\(cf. Fig. \ref{fig:composed-classifier})}}
& Name of ComposedRosNodeClassifier & Trace 1.1 \& 1.2\\
\cline{2-3}
& Stereotype of ComposedRosNodeClassifier & Trace 1.1 \& 1.2\\
\cline{2-3}
& Namespace Name & Trace 3\\
\cline{2-3}
& Namespace Stereotype & Trace 3\\
\cline{2-3}
& Remapped Topic Name & Traces 4.1 \& 4.2\\
\cline{2-3}
& Name of RosNodePart & Traces 1.2 \& 2\\
\cline{2-3}
& Stereotype of RosNodePart & Traces 1.2 \& 2\\
\cline{2-3}
& Type of RosNodePart & Traces 1.2 \& 2\\
\bottomrule
\end{tabular}
\end{table}

\subsection{Evaluation Results}
In \textbf{Case Study~I}, we deliberately exclude launch artifacts. Reconstruction therefore relies only on source-level node class definitions, yielding six \textit{AtomicRosNodeClassifierDiagram} instances and no \textit{ComposedRosNodeClassifierDiagram} (cf. Fig.~\ref{fig:atomic-classifier} and Fig.~\ref{fig:composed-classifier}).

Table~\ref{tab:eval-results-c1} reports precision, recall, and F1-score of $1.0$ for all evaluated metric elements (cf. Table~\ref{tab:metrics-impl}). This is expected: the setting is deterministic and purely atomic, and the code contains only lightweight functionality, keeping structural cues explicit and easy to extract. This enables a direct syntactic-to-architectural mapping from source constructs. The few-shot examples (cf. Section~\ref{section:recoveryarch}) are repository-independent and only reinforce the blueprint schema. Without hierarchical composition or launch-induced transformations, the task reduces to structural extraction, for which conventional static analysis would achieve comparable results.

\begin{table}[ht]
\centering
\footnotesize
\setlength{\tabcolsep}{4pt}
\caption{Evaluation Results for Case Study I}
\label{tab:eval-results-c1}
\begin{threeparttable}
\begin{tabular}{m{0.6cm} m{3.2cm} m{1.1cm} m{1.1cm} m{1.1cm}}
\toprule
 & \textbf{Metric Element} & \textbf{Precision} & \textbf{Recall} & \textbf{F1}\\
\midrule
\multirow{6}{*}[+2.5em]{\verticalcell[3.5cm]{\textbf{AtomicClassifier-}\\\textbf{Diagram}}}
& ARC Name
& \textit{avg} = \textbf{1.0}
& \textit{avg} = \textbf{1.0}
& \textit{avg} = \textbf{1.0}\\
\cline{2-5}
& ARC Stereotype
& \textit{avg} = \textbf{1.0}
& \textit{avg} = \textbf{1.0}
& \textit{avg} = \textbf{1.0}\\
\cline{2-5}
& Message Type
& \textit{avg} = \textbf{1.0}
& \textit{avg} = \textbf{1.0}
& \textit{avg} = \textbf{1.0}\\
\cline{2-5}
& Callback Function Name
& \textit{avg} = \textbf{1.0}
& \textit{avg} = \textbf{1.0}
& \textit{avg} = \textbf{1.0}\\ 
\cline{2-5}
& Service Type
& \textit{avg} = \textbf{1.0}
& \textit{avg} = \textbf{1.0}
& \textit{avg} = \textbf{1.0}\\
\cline{2-5}
& Service Function Name
& \textit{avg} = \textbf{1.0}
& \textit{avg} = \textbf{1.0}
& \textit{avg} = \textbf{1.0}\\
\midrule
& \textbf{Average across Elements} & \textbf{1.0} & \textbf{1.0} & \textbf{1.0} \\
\bottomrule
\end{tabular}
\begin{tablenotes}[flushleft]
\footnotesize
\item \textbf{Notes:} ARC = AtomicRosNodeClassifier; \emph{avg} denotes the arithmetic mean of the six vector entries, where each entry represents the precision, recall, and F1-score of one \textit{AtomicRosNodeClassifier} instance.
\end{tablenotes}
\end{threeparttable}
\end{table}

In \textbf{Case Study~II}, launch files introduce hierarchical instantiation and namespace scoping, requiring cross-artifact integration. Six node classes instantiated across three launch files form three composed subsystems. The code is intentionally lightweight and contains little functional logic, which keeps structural cues explicit. Accordingly, atomic recovery remains perfect (P = R = F1 = 1.0), confirming deterministic extraction of definition-level elements. At the composed level, precision stays high (0.88) while recall drops (0.75, F1 = 0.81), suggesting that the blueprint constrains the solution space and avoids spurious elements, whereas missed results stem from incomplete reconstruction of integration relations, in particular namespace scoping and composition semantics.

In \textbf{Case Study~III}, repository complexity increases substantially in an industrial-scale ROS~2 codebase with substantial functional logic. This additional functionality and abstraction obscures structural cues and contributes to the performance drop. Atomic performance decreases to P = 0.78, R = 0.55 (F1 = 0.64). In Autoware, node implementations often rely on multi-level inheritance rather than direct extension of \texttt{rclcpp::Node}, and communication endpoints are frequently wrapped in helper classes (e.g., \texttt{DebugPublisher}), masking explicit \texttt{rclcpp} API usage. At the composed level, precision remains high (1.0) but recall declines further (0.35, F1 = 0.49), indicating structurally valid yet incomplete subsystem reconstruction. Overall, the blueprint preserves architectural validity (precision), while recall becomes increasingly sensitive to implicit integration semantics in large-scale systems.

Compared to related automated UML- and domain-model recovery approaches, the results remain competitive. Yang and Sahraoui~\cite{yang2022towardsautouml} report end-to-end F1-scores below 0.50 even under controlled natural-language settings. Chen et al.~\cite{Chen2023AutomatedDomainModeling} report best-case F1-scores of 0.76 (classes), 0.61 (attributes), and 0.34 (relationships) when using GPT-4 for automated domain modeling. In contrast, our approach achieves excellent atomic recovery in structurally explicit repositories and comparable subsystem-level F1 under substantially more heterogeneous ROS~2 conditions.

Across the three case studies, a consistent trend emerges: precision stays high, confirming the effectiveness of blueprint-guided constraints, while recall decreases as implementations become more complex and logic-heavy and as launch artifacts introduce implicit integration semantics. This identifies integration-phase reconstruction---rather than structural validity---as the primary remaining challenge.

\begin{table}[t]
\centering
\footnotesize
\setlength{\tabcolsep}{4pt}
\renewcommand{\arraystretch}{1.05}
\caption{Evaluation Results for Case Study II and III}
\label{tab:eval-results-c2-c3}
\begin{threeparttable}
\begin{tabular}{m{0.7cm} m{0.6cm} m{2.0cm} m{2.0cm} m{2.0cm}}
\toprule
& \textbf{Level} & \textbf{Precision} & \textbf{Recall} & \textbf{F1}\\
\midrule
\multirow{2}{*}[+0.7em]{\verticalcell[1.35cm]{\textbf{Case Study II}}}
& ACD
& [1.0,~1.0,~1.0], \textit{avg} = \textbf{1.0}
& [1.0,~1.0,~1.0], \textit{avg} = \textbf{1.0}
& [1.0,~1.0,~1.0], \textit{avg} = \textbf{1.0}\\
\cline{2-5}
& CCD
& [1.0, 0.60, 1.0], \textit{avg} = \textbf{0.88}
& [0.80, 0.43, 1.0], \textit{avg} = \textbf{0.75}
& [0.89, 0.50, 1.0], \textit{avg} = \textbf{0.81}\\
\midrule
\multirow{2}{*}[+0.7em]{\verticalcell[1.35cm]{\textbf{Case Study III}}}
& ACD
& [0.80,~0.75,~0.80], \textit{avg} = \textbf{0.78}
& [0.57,~0.67,~0.40], \textit{avg} = \textbf{0.55}
& [0.67,~0.71,~0.53], \textit{avg} = \textbf{0.64}\\
\cline{2-5}
& CCD
& [1.00,~1.00], \textit{avg} = \textbf{1.00}
& [0.20,~0.49], \textit{avg} = \textbf{0.35}
& [0.33,~0.65], \textit{avg} = \textbf{0.40}\\
\bottomrule
\end{tabular}
\begin{tablenotes}[flushleft]
\footnotesize
\item \textbf{Notes:} ACD = AtomicClassifierDiagram; CCD = ComposedClassifierDiagram (cf. Fig. \ref{fig:atomic-classifier} and Fig. \ref{fig:composed-classifier}); Vector entries report per-instance scores; \emph{avg} denotes the arithmetic mean of entries in a vector.
\end{tablenotes}
\end{threeparttable}
\end{table}

\section{Conclusion, Limitations and Future Work}
This paper presented a blueprint-guided hybrid approach to recover hierarchical structural architectures of ROS~2 systems from code and configuration artifacts. We introduced a UML-based modeling concept as an explicit target representation and a recovery pipeline that combines deterministic extraction with blueprint-constrained LLM-based synthesis and validation.

Our evaluation across three ROS~2 repositories shows high precision across abstraction levels, indicating that blueprint constraints effectively enforce structural validity. At the same time, recall decreases for higher abstraction levels in larger repositories, because subsystem composition and integration semantics are often dispersed and implicitly encoded in configuration artifacts. This highlights integration-phase reconstruction as the main remaining challenge for large-scale and industrial systems.

Future work includes consolidating the modeling concept into a fully formal UML profile with an explicit metamodel and well-formedness constraints, extending support for complex launch constructs, and integrating the approach into CI pipelines for continuous documentation generation and architectural drift detection. Beyond static structural decomposition, future extensions should incorporate behavioral intent, quality attributes, and runtime-induced effects, complemented by bounded human-in-the-loop clarification for ambiguous integration semantics. Finally, establishing publicly available ROS~2 benchmarks with reference architectures and shared metrics would improve reproducibility and cross-approach comparability.

\printbibliography

@article{Winiarski_2023,
   title={MeROS: SysML-Based Metamodel for ROS-Based Systems},
   volume={11},
   ISSN={2169-3536},
   journal={IEEE Access},
   publisher={Institute of Electrical and Electronics Engineers (IEEE)},
   author={Winiarski, Tomasz},
   year={2023},
   pages={82802–82815} }

@article{10.1145/3763169,
author = {Canelas, Paulo and Schmerl, Bradley and Fonseca, Alcides and Timperley, Christopher S.},
title = {ROSpec: A Domain-Specific Language for ROS-Based Robot Software},
year = {2025},
issue_date = {October 2025},
publisher = {Association for Computing Machinery},
address = {New York, NY, USA},
volume = {9},
number = {OOPSLA2},
journal = {Proc. ACM Program. Lang.},
month = oct,
articleno = {391},
numpages = {29}
}

@INPROCEEDINGS{7416545,
  author={Kumar, Pranav Srinivas and Emfinger, William and Kulkarni, Amogh and Karsai, Gabor and Watkins, Dexter and Gasser, Benjamin and Ridgewell, Cameron and Anilkumar, Amrutur},
  booktitle={2015 International Symposium on Rapid System Prototyping (RSP)}, 
  title={ROSMOD: a toolsuite for modeling, generating, deploying, and managing distributed real-time component-based software using ROS}, 
  year={2015},
  volume={},
  number={},
  pages={39-45}
}

@INPROCEEDINGS{7926539,
  author={Bardaro, Gianluca and Matteucci, Matteo},
  booktitle={2017 First IEEE International Conference on Robotic Computing (IRC)}, 
  title={Using AADL to Model and Develop ROS-Based Robotic Application}, 
  year={2017},
  volume={},
  number={},
  pages={204-207},
}

@inproceedings{Wanninger2021,
  author = {Wanninger, Constantin and Rossi, Sebastian and Schörner, Martin and Hoffmann, Alwin and Poeppel, Alexander and Eymüller, Christian and Reif, Wolfgang},
  title = {{ROSSi} a graphical programming interface for {ROS 2}},
  booktitle = {2021 21st International Conference on Control, Automation and Systems (ICCAS)},
  year = {2021},
  pages = {255-262},
  publisher = {IEEE},
  address = {Piscataway, NJ},
}

@misc{hatahet2025generatingsoftwarearchitecturedescription,
      title={Generating Software Architecture Description from Source Code using Reverse Engineering and Large Language Model}, 
      author={Ahmad Hatahet and Christoph Knieke and Andreas Rausch},
      year={2025},
      eprint={2511.05165},
      archivePrefix={arXiv},
      primaryClass={cs.SE},
}

@inproceedings{fuchs2025enabling,
  author = {Fuchß, D. and Liu, H. and Hey, T. and Keim, J. and Koziolek, A.},
  title = {Enabling Architecture Traceability by {LLM}-Based Architecture Component Name Extraction},
  booktitle = {2025 IEEE 22nd International Conference on Software Architecture (ICSA)},
  pages = {1--12},
  publisher = {IEEE},
  year = {2025},
}

@article{brown2013software,
  title={Software architecture for developers},
  author={Brown, Simon},
  journal={Coding the Architecture},
  year={2013}
}

@article{zhao2026softwarear,
  title={Software Architecture Recovery Augmented With Semantics},
  author={Wenting Zhao and Wuxia Jin and Yiran Zhang and Mi-Qian Fan and Haijun Wang and Li Li and Yang Liu and Ting Liu},
  journal={IEEE Transactions on Software Engineering},
  year={2026},
  volume={52},
  pages={338-359},
}

@online{crewai2024introduction,
  author = {{CrewAI Team}},
  title = {Introduction to CrewAI},
  year = {2024},
  url = {https://docs.crewai.com/en/introduction},
  note = {Accessed: 2026-02-17}
}

@article{koschke2005rekonstruktion,
  author  = {Koschke, R.},
  title   = {Rekonstruktion von Software-Architekturen},
  journal = {Informatik Forsch. Entw.},
  volume  = {19},
  pages   = {127--140},
  year    = {2005},
}

@inproceedings{yang2022towardsautouml,
author = {Yang, Song and Sahraoui, Houari},
title = {Towards automatically extracting UML class diagrams from natural language specifications},
year = {2022},
publisher = {Association for Computing Machinery},
address = {New York, NY, USA},
booktitle = {Proceedings of the 25th International Conference on Model Driven Engineering Languages and Systems: Companion Proceedings},
pages = {396–403},
numpages = {8},
series = {MODELS '22}
}

@inproceedings{10.1145/3550356.3561583,
author = {Singh, Prabhsimran and Boubekeur, Younes and Mussbacher, Gunter},
title = {Detecting mistakes in a domain model},
year = {2022},
publisher = {Association for Computing Machinery},
address = {New York, NY, USA},
pages = {257–266},
numpages = {10},
series = {MODELS '22}
}

@inproceedings{10.1145/3417990.3418741,
author = {Boubekeur, Younes and Mussbacher, Gunter and McIntosh, Shane},
title = {Automatic assessment of students' software models using a simple heuristic and machine learning},
year = {2020},
publisher = {Association for Computing Machinery},
address = {New York, NY, USA},
booktitle = {Proceedings of the 23rd ACM/IEEE International Conference on Model Driven Engineering Languages and Systems: Companion Proceedings},
articleno = {20},
numpages = {10},
series = {MODELS '20}
}

@article{SONG2024123927,
title = {A deep learning-based approach to similarity calculation for UML use case models},
journal = {Expert Systems with Applications},
volume = {251},
pages = {123927},
year = {2024},
author = {Shizhe Song and Ye Wang and Xiaoyang Wang and Chengyi Lin and Kun Hu},
}

@article{yuan2020structural,
  author    = {Yuan, Zhen and Yan, Lian and Ma, Zhi},
  title     = {Structural similarity measure between {UML} class diagrams based on {UCG}},
  journal   = {Requirements Engineering},
  year      = {2020},
  volume    = {25},
  pages     = {213--229},
  number    = {2},
  publisher = {Springer}
}

@inproceedings{Chen2023AutomatedDomainModeling,
  author    = {Kua Chen and Yujing Yang and Boqi Chen and
               Jos{\'e} Antonio Hern{\'a}ndez L{\'o}pez and
               Gunter Mussbacher and D{\'a}niel Varr{\'o}},
  title     = {Automated Domain Modeling with Large Language Models: A Comparative Study},
  booktitle = {Proceedings of the 2023 ACM/IEEE 26th International Conference on Model Driven Engineering Languages and Systems (MODELS)},
  year      = {2023},
  pages     = {162--172},
  publisher = {IEEE},
  doi       = {10.1109/MODELS58315.2023.00037}
}

@book{starke2019arc42,
  title={arc42 by Example: Software architecture documentation in practice},
  author={Starke, G. and Simons, M. and Z{\"o}rner, S. and M{\"u}ller, R.D.},
  isbn={9781839219269},
  url={https://books.google.de/books?id=1hW1DwAAQBAJ},
  year={2019},
  publisher={Packt Publishing}
}

@ARTICLE{schmidt,
  author={Schmidt, D.C.},
  journal={Computer}, 
  title={Guest Editor's Introduction: Model-Driven Engineering}, 
  year={2006},
  volume={39},
  number={2},
  pages={25-31},
  doi={10.1109/MC.2006.58}
}

@ARTICLE{semarch,
  author={Zhao, Wenting and Jin, Wuxia and Zhang, Yiran and Fan, Ming and Wang, Haijun and Li, Li and Liu, Yang and Liu, Ting},
  journal={IEEE Transactions on Software Engineering}, 
  title={Software Architecture Recovery Augmented With Semantics}, 
  year={2026},
  volume={52},
  number={1},
  pages={338-359},
  doi={10.1109/TSE.2025.3620670}
}

\vspace{12pt}

\end{document}